\documentclass[aps,prl,showpacs,amsmath,amssymb,amsfonts,lengthcheck,twocolumn,longbibliography,superscriptaddress]{revtex4-2}
\usepackage{epsfig,graphicx,graphics,amsmath,amssymb,float}
\usepackage[T1]{fontenc}
\usepackage[latin9]{inputenc}
\usepackage{amsmath}
\usepackage{amssymb}
\usepackage{xcolor}
\usepackage{amscd}
\usepackage{bm}
\usepackage{bbold}
\usepackage{psfrag}
\usepackage{mathrsfs}
\usepackage{float}

\usepackage{bbm} 

\usepackage{chemformula}
\usepackage[caption=false]{subfig}
\usepackage{subfig}
\usepackage{color}
\usepackage[bookmarks=true,colorlinks,linkcolor=blue,urlcolor=blue,citecolor=blue]{hyperref}

\usepackage{dutchcal}
\DeclareMathAlphabet{\matholdcal}{OMS}{cmsy}{m}{n}

\DeclareMathOperator{\sgn}{sgn} 

\DeclareMathOperator{\Tr}{Tr} 
\DeclareMathOperator{\Pf}{Pf}

\newcommand{\be}{\begin{equation}}
\newcommand{\ee}{\end{equation}}
\newcommand{\bea}{\begin{eqnarray}}
\newcommand{\eea}{\end{eqnarray}}

\newcommand{\mc}{\matholdcal}

\newcommand{\te}{\text}
\begin{document}

\title{Emergence of $X$ states in a quantum impurity model}%SD

\author{Moallison F. Cavalcante}
\email{moallison@umbc.edu}
\affiliation{Department of Physics, University of Maryland, Baltimore County, Baltimore, MD 21250, USA}
\affiliation{Quantum Science Institute, University of Maryland, Baltimore County, Baltimore, MD 21250, USA}
\affiliation{
Gleb Wataghin Physics Institute, The University of Campinas, 13083-859, Campinas, S\~{a}o Paulo, Brazil}
\author{ Marcus V. S. Bonan\c{c}a}
\affiliation{
Gleb Wataghin Physics Institute, The University of Campinas, 13083-859, Campinas, S\~{a}o Paulo, Brazil}
\author{Eduardo Miranda}
\affiliation{
Gleb Wataghin Physics Institute, The University of Campinas, 13083-859, Campinas, S\~{a}o Paulo, Brazil}
\author{Sebastian Deffner}
\affiliation{Department of Physics, University of Maryland, Baltimore County, Baltimore, MD 21250, USA}
\affiliation{Quantum Science Institute, University of Maryland, Baltimore County, Baltimore, MD 21250, USA}
\affiliation{National Quantum Laboratory, College Park, MD 20740, USA}
\date{\today}

\begin{abstract}
 
In the present work, we demonstrate the emergence of $X$-states in the long-time response of a locally perturbed many-body quantum impurity model. The emergence of the double-qubit state is heralded by the lack of decay of the response function as well as the out-of-time order correlator, signifying the trapping of excitations and hence information in edge modes. Surprisingly, after carrying out a quantum information theory characterization, we show that such states exhibit genuine quantum correlations. 
\end{abstract}

%\pacs{pacs}

\maketitle

\paragraph*{Introduction}
\label{introduction}

Recent years have seen the rapid development of more and more advanced quantum computing platforms. Leading technologies are based on a variety of physical platforms \cite{Sanders2017}, such as ion traps \cite{BruzewiczAPR2019}, neutral atoms \cite{Wintersperger2023}, entangled photons \cite{Slussarenko2019APR}, or superconducting circuits \cite{Ezratty2023EPJA}. In the present work, by demonstrating the emergence of an entangled \emph{double-qubit} $X$-state in an interacting quantum many-body system, we provide an alternative way of storing and processing quantum information using localized edge modes. 

%In the present work, we demonstrate the emergence of an entangled \emph{double-qubit} $X$-state in an interacting quantum many-body model. Notably, this state is realized in the nonequilibrium stationary state of the local edge modes. 

%A uniquely distinct approach is based on leveraging the topological properties of complex quantum many-body systems and the resulting anyonic statistics of the quasiparticle excitations \cite{LahtinenSPPhys2017}.

%In the present work, we demonstrate the emergence of a topological $X$-state in an impurity model. Notably, this \emph{topological double-qubit} is realized in the stationary state of the local edge modes, and hence it is a consequence of the topology of the quantum many-body system. In contrast to more conventional topological qubits, however, our $X$-state is not characterized by anyonic statistics, but rather by the nonequilibrium response of the perturbed system.

Understanding the nonequilibrium dynamics of closed, one-dimensional quantum many-body systems is a challenging problem. Due to the inherent complexity that emerges from the interaction between the many degrees of freedom, general descriptions accounting for universal dynamical phenomena is a rather involved problem \cite{Whaley2015NJP,SorianiPRL2022,CavalcantePRB2024}. Therefore, it is paramount to investigate models that exhibit analytical, or partially analytical, solutions, where more controllable and in-depth studies can be developed. In this direction, quantum impurity models have attracted significant attention in the literature \cite{PRL_2013_Vasseur,Vasseur2014,PRX_quench_Majo_2014,PRB_trans_LL_2015,PRB_main_ref_2016,PRL_Bertini_2016,OTOC_Kondo_PRB_2017,PRB_spin_relax_2021,PRB_count_edge_2023,fastscramblingboundary_2024}.

\begin{figure}
    \centering
    \includegraphics[scale=0.22]{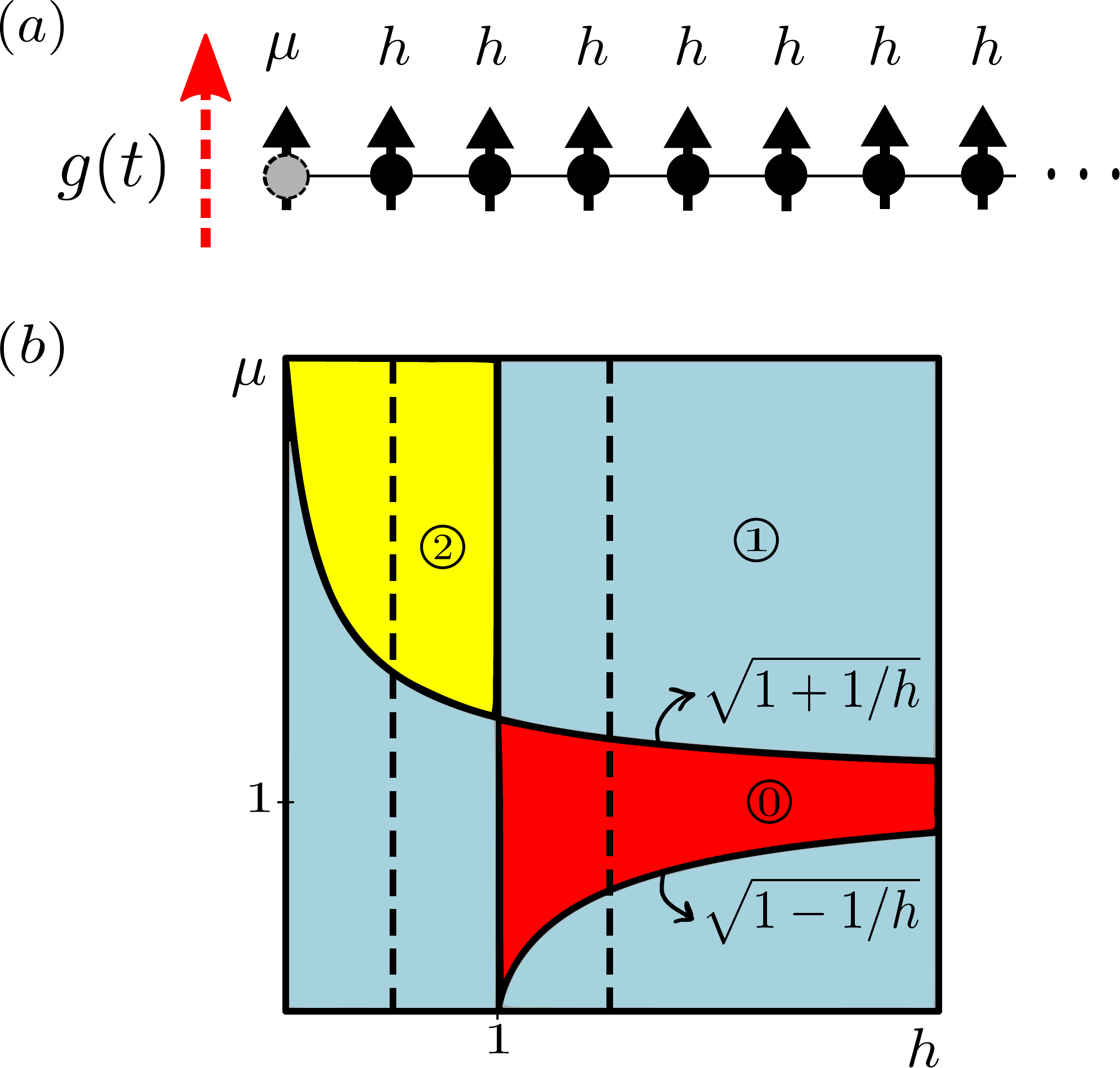}
    \caption{(a) Sketch of the model in Eq.~(\ref{main_Hamil}). The impurity ($\mu)$ is described by the gray ball. The large red arrow represents the finite-time control $g(t)$ acting at the impurity site. (b) $(h,\mu)$-phase diagram \cite{PRB_main_ref_2016}. The different colored regions highlight the number of localized modes close to the edge $j=1$: zero in the red region, one in the light blue region, and two in the yellow region. The two dashed black lines indicate two values of the control parameter $h$, namely, $h=1/2$ and $3/2$, which we will use later on in order to cover all the boundary phases.
    }    
    \label{fig_system}
\end{figure}

Here, we study the dynamical response of a quantum impurity model, namely, the transverse-field quantum Ising chain with an impurity at its edge \cite{PRB_main_ref_2016}. Such an impurity gives rise to a rich phase diagram, where new boundary phases, with up to two localized edge states, emerge, see Fig. \ref{fig_system}(b) below. We show that these edge modes can be accessed through the local control of the impurity and that they hinder the perfect spread of energy and information throughout the chain. In other words, for a suitable finite-time control protocol, excitations can be introduced in the chain through the impurity and a part of them gets trapped in the localized modes. After relaxation, the state describing the remaining excitations can take the form of two entangled qubits, also known as an $X$-state \cite{Rau_2009}. This opens the possibility of using this setup for information-processing purposes.

\paragraph*{Quantum impurity model}
\label{sec_model}

We start by describing the system. We consider an interacting quantum system, namely, a spin-$1/2$ chain with open boundary conditions described by the following Hamiltonian,
\begin{eqnarray}   
 H_{0}&=&-J\sum_{j=1}^{N-1}\sigma^x_{j}\sigma^x_{j+1} - \:Jh\sum_{j=2}^{N}\sigma^z_{j}-Jh\mu \sigma^z_1,
\label{main_Hamil}    
\end{eqnarray}
where $\sigma^{a=x,y,z}_j$ are the Pauli matrices (two times the spin-$1/2$ operators) at the site $j$, $J\equiv 1$ is the exchange coupling, $h>0$ is the external field magnitude and $\mu\neq 1$ represents the impurity at the edge. Typically, we will assume that $N\to \infty$. 

In the fermionic representation, that is, applying a standard Jordan-Wigner transformation \cite{LIEB1961407,sachdev1999quantum}, the model in Eq. (\ref{main_Hamil}) becomes (see Suppl. Mat.~\cite{SM}),
\begin{equation}
    H_0=\sum_{ij}\left[c^{\dagger}_iA_{ij}c_j+\frac{1}{2}\left(c^{\dagger}_iB_{ij}c^{\dagger}_j +\text{H.c.}\right)\right],
    \label{imp_Hamil_2} 
\end{equation}
where $A_{ij}=2h\left[(1-\mu)\delta_{i1}-1\right]\delta_{ij}+ \left(\delta_{i,j+1}+\delta_{i+1,j}\right)=A_{ji}$ and $B_{ij}=(\delta_{i+1,j}-\delta_{i,j+1})=-B_{ji}$. In this fermionic form, we see that the impurity plays the role of a local potential at the edge of a one-dimensional spinless $p$-wave superconductor \cite{Kitaev_2001,Alicea_2012,CMoore_PRB_2018}.

This model was introduced and analytically solved in Ref.~\cite{PRB_main_ref_2016}. In Suppl. Mat. \cite{SM}, 
we provide the main details of this solution. The authors of Ref.~\cite{PRB_main_ref_2016} have shown that the presence of the edge impurity, $\mu\neq 1$, produces a new localized (nontopological) mode at the edge $j=1$, beyond the well-studied topological edge mode found by Kitaev in the impurity-free case, $\mu=1$~\cite{Kitaev_2001}.  Although slightly different from the transverse field quantum Ising chain, the model in Eq. (\ref{main_Hamil}) still undergoes a quantum phase transition when (for $J=1$) $h$ is changed and crosses the critical value $h_c \equiv 1$ \cite{PRB_main_ref_2016}. For values $h<h_{c}$ and $h>h_c$, the system is found in the ferromagnetic and paramagnetic phases, respectively. 

We highlight that, different from the delocalized (bulk) modes, the two localized edge modes, which hereafter we will denote by $\gamma_{1,2}$,  
\textit{only} appear in specific regions of the $(h,\mu)$-phase diagram (see Fig. \ref{fig_system}(b) below). Notice that $\gamma_{1,2}$ are two complex fermionic modes, $\gamma^{\dagger}_i\neq \gamma_i$. Along the line $\mu=1$, we recover a physics reminiscent of the Kitaev chain \cite{Kitaev_2001}: for $h<1$, the mode $\gamma_1$ exists (topological phase), while for $h>1$ only the delocalized modes are present (trivial phase). The mode $\gamma_1$ is made up of two real (Majorana) modes, $\gamma_1=(\alpha+i\beta)/2$, where $\alpha$ is well localized around $j=1$ and $\beta$ around $j=N$. For a semi-infinite chain $N\to\infty$, these real fermions are practically decoupled from each other, and thus we have two isolated Majorana fermions. For $\mu\neq 1$, the impurity effects appear. However, the mode $\gamma_1$ is still present for all $\mu$ in the whole ferromagnetic side, reflecting its robustness against local perturbations. 
Concerning the mode $\gamma_2$, it can appear in two different regions, depending on the values of $(h,\mu)$. These two regions are: $h>1$ and $|\mu|<\sqrt{1-1/h}$, and, $\forall \:h$ and $|\mu|>\sqrt{1+1/h}$. 

Figure~\ref{fig_system}(b) summarizes this information as follows: in the light blue regions, we only have one localized mode, either $\gamma_1$ or $\gamma_2$, while in the yellow region we have both localized modes. Finally, in the red region, only the delocalized modes are present. Furthermore, it is important to stress once again that the $\gamma_{1,2}$ modes have very different physical origins. The $\gamma_2$ mode is a direct consequence of the impurity $\mu\neq 1$. Putting in context, this mode represents the partially separated Andreev bound state (strongly overlapping here) 
widely discussed in~\cite{CMoore_PRB_2018}, thus it is not topologically protected. On the other hand, the $\gamma_1$ mode has a topological origin and hence it is topologically protected \cite{Kitaev_2001,Alicea_2012,PRB_Wade_2013}. The real-time dynamics of such localized edge states has been experimentally investigated in \cite{Meier2016}.
 
\begin{figure*}
    \centering
    \includegraphics[scale=0.5]{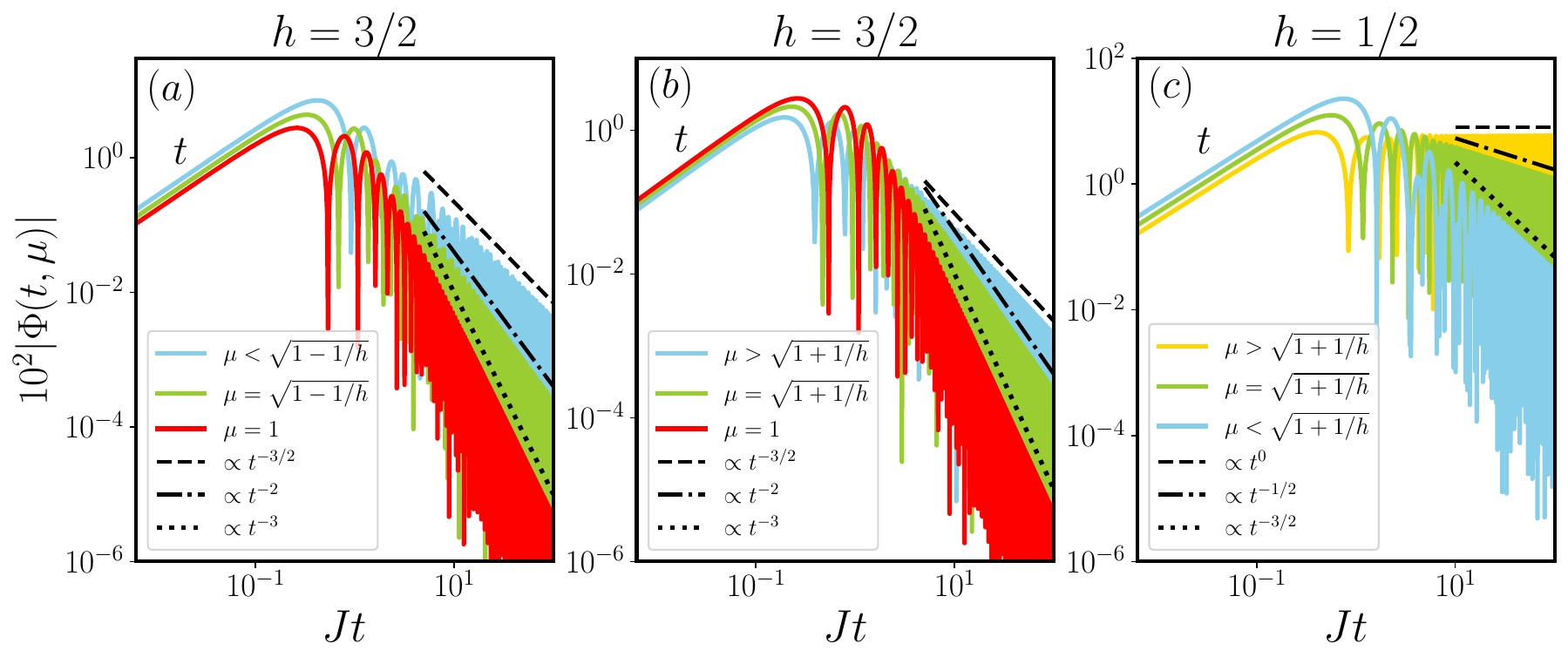}
    \caption{Response function (\ref{respon_fun_2}) for different $\mu$ and for two values of $h$. For very short times $t \ll J^{-1}$, we can see the general $t$ growth. In (a), (b) and  (c), we go through the different boundary phases:  (a) and (b) correspond to the paramagnetic side while (c) to the ferromagnetic one (see Fig. \ref{fig_system}(b)). The dashed, dashed-dotted and dotted black lines represent the power-law decays obtained from stationary phase approximation (see Suppl. Mat.~\cite{SM} for details).}
    \label{fig_resp_func}
\end{figure*}

In its diagonal form, the Hamiltonian in Eq. (\ref{main_Hamil}) reads,
\begin{eqnarray}
H_0&=&\sum_{\kappa}\Gamma_{\kappa}\gamma^{\dagger}_{\kappa}\gamma_{\kappa}.
    \label{main_Hamil_2}
\end{eqnarray}
In the latter equation, summation over $\kappa$ includes both localized, $\gamma_{1,2}$, and delocalized, $\gamma_{k}$, fermionic modes. 
The ground state $|\mathbf 0\rangle$ of the Hamiltonian (\ref{main_Hamil}) is the fermionic vacuum of $\gamma_{\kappa}$, thus $\gamma_{\kappa}|\mathbf 0\rangle=0$ for all $\kappa$. In Suppl. Mat. \cite{SM}, we provide explicit expressions for the dispersions $\Gamma_{\kappa}$.

\paragraph*{Local control and response function \label{sec_resp_fun}}

We now turn to the response of the system to finite-time control localized at $j=1$. Assuming that the chain is initialized in the ground state $|\mathbf 0\rangle$ (which is a zero-temperature equilibrium state) for arbitrary values of $(h,\mu)$, we apply an impulse $g(t)=-g_{0}\delta(t)\sigma_{1}^{z}$ at the impurity (where $\delta(\cdot)$ denotes the Dirac delta function). 

This perturbation drives the system out of equilibrium and injects energy. An experimentally accessible quantity \cite{Cheneau_2012} is the response function,
\begin{eqnarray}
    \Phi(t,\mu)&=&\frac{i}{2^3}\langle [\sigma^z_1(t),\sigma^z_1(0)] \rangle_0,
    \label{respon_fun}
\end{eqnarray}
where $\sigma^z_1(t)=e^{iH_0t}\sigma^z_1e^{-iH_0t}$ is the operator $\sigma^z_1$ in the Heisenberg representation \cite{coleman2015introduction} and $\langle\cdots\rangle_0\equiv\langle\mathbf 0|\cdots |\mathbf 0\rangle$. For a sufficiently small value of $g_{0}$, the impulse is weak and the response function (\ref{respon_fun}) describes how $\langle \sigma_{1}^{z}(t)\rangle$ evolves \cite{Kubo}. In the following, we will show that the local monitoring of the system at the impurity provides important information about the nonequilibrium dynamics of the considered model in its different boundary phases.

Firstly, we consider the short-time behavior of $\Phi(t,\mu)$. For very short times $t\ll J^{-1}$, we expand the operator $\sigma^z_1(t)$ in powers of $Jt$ to obtain $\Phi(t\ll J^{-1},\mu) \propto t$, regardless of the point in the $(h,\mu)$-phase diagram.
In other words, the short-time behavior of the response function does not distinguish the different boundary phases. 

To obtain intermediate and long-time behaviors of $\Phi(t,\mu)$, we use the diagonal form (\ref{main_Hamil_2}) of the Hamiltonian $H_0$. This allows us to perform the calculation of the expectation value in Eq. (\ref{respon_fun}), yielding
\begin{eqnarray}
    \Phi(t,\mu)&=&\sum_{\kappa,\kappa'}f_{\kappa\kappa'}\sin[(\Gamma_{\kappa}+\Gamma_{\kappa'})t].
    \label{respon_fun_2}
\end{eqnarray}
The amplitude $f_{\kappa\kappa'}$ is the probability of finding the system in the two-particle state $|\kappa \kappa'\rangle=\gamma^{\dagger}_{\kappa}\gamma^{\dagger}_{\kappa'}|\mathbf 0\rangle$ after the previously described perturbation. It reads, $f_{\kappa\kappa'}=v_{\kappa}u_{\kappa'}(v_{\kappa}u_{\kappa'}-v_{\kappa'}u_{\kappa})$, with $u_{\kappa}= u_{k}$ and $v_{\kappa}= v_{k}$ for the delocalized modes and $u_{\kappa}=u^{(\ell)}$ and $v_{\kappa}=v^{(\ell)}$ for $\gamma_{\ell=1,2}$. These are wave functions at the edge, $j=1$, for the different modes, see Suppl. Mat. \cite{SM} to find explicit expressions.

The colored curves in Fig.~\ref{fig_resp_func} show Eq.~(\ref{respon_fun_2}) for different values of $\mu$ and for the values of $h$ corresponding to the dashed black lines in Fig.~\ref{fig_system}(b). Figures~\ref{fig_resp_func}(a, b) depict the results for the paramagnetic phase, $h>1$. After an initially linear growth, around $t\approx J^{-1}$, the response function starts to decrease as an oscillatory function with a power-law envelope. 
In the regions with only one localized mode (the $\gamma_2$ mode), a slower relaxation is observed for $\Phi(t,\mu)$, that is, it tends to zero according to $t^{-3/2}$. 
Figures~\ref{fig_resp_func}(a, b) also show a faster decay of $\Phi(t,\mu)$ along the boundaries $\mu=\sqrt{1\pm 1/h}$, $\Phi(t,\mu)\sim t^{-2}$, although the $\gamma_2$ mode is absent. Deep inside the red region of Fig.~\ref{fig_system}(b) (no edge modes), we have a $t^{-3}$ scaling for $\Phi(t,\mu)$.

Figure~\ref{fig_resp_func}(c) shows the results for the ferromagnetic phase, $h<1$.  In the region where only the mode $\gamma_1$ is present ($\mu<\sqrt{1+1/h}$), the scaling behavior $t^{-3/2}$ is exactly the same as that observed in the paramagnetic region (when $\gamma_2$ exists). Similarly to the paramagnetic side, the power-law decay changes at the boundary between two different phases. Along $\mu=\sqrt{1+1/h}$, a much slower decay can be observed, that is, $\Phi(t\gg J^{-1},\mu)\sim t^{-1/2}$. Again, the impurity effects can be detected by the response function even when the mode $\gamma_2$ (which is a direct consequence of it) is absent. 

Deep inside the yellow region of Fig.~\ref{fig_system}(b), both modes $\gamma_{1,2}$ exist and contribute to the response function. In Fig.~\ref{fig_resp_func}(c), we notice their drastic effect on $\Phi(t\gg J^{-1},\mu)$: it no longer decays to zero but instead oscillates indefinitely.
In fact, we obtain
\begin{eqnarray}
\Phi(t\gg J^{-1},\mu)&\propto&\sin \Gamma^{(2)}t+\:\mc O(t^{-3/2}),
\label{Respon_func_two_mode}
\end{eqnarray}
where $\Gamma^{(2)}$ is the energy of the mode $\gamma_2$ (since the mode $\gamma_1$ has zero energy in the large $N$ limit). This coherent oscillation in the long-time limit appears after the delocalized modes undergo relaxation, and so the dynamics is effectively governed only by the two edge modes. In summary, the response function (\ref{respon_fun}) already shows that local monitoring of the system response detects the partial trapping of excitations injected by the local control.

\begin{figure*}
    \centering
    \includegraphics[scale=0.5]{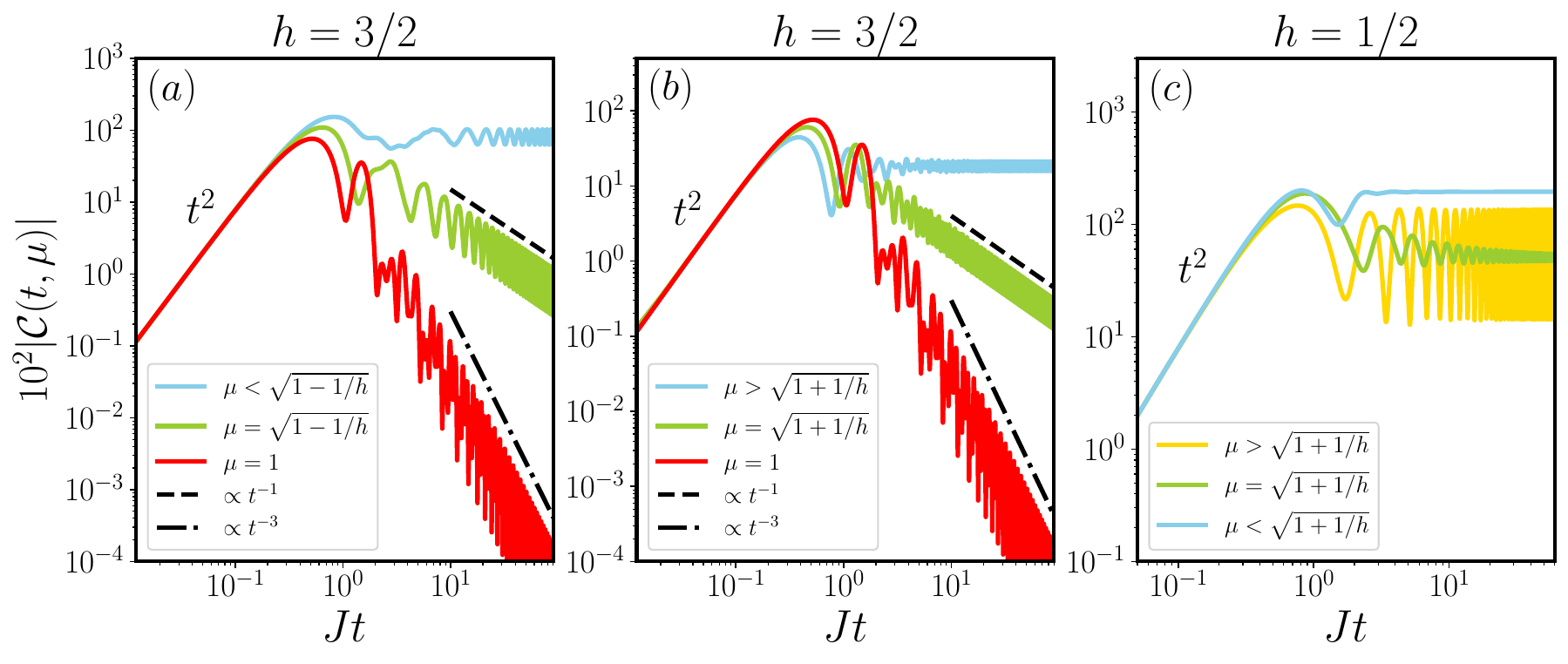}
    \caption{OTOC in Eq. (\ref{OTOC_main}) for different values of $\mu$ and for two values of $h$. The general short-time growth $t^2$ for $t\ll J^{-1}$ is observed. Once again, in (a), (b) and (c) we go through the different boundary phases shown in Fig. \ref{fig_system}(b). In (a) and (b), we show results on the paramagnetic side, while in (c) the ferromagnetic behaviors are shown. The dashed and dashed-dotted black lines in (a) and (b) represent two power-law decays (see Suppl. Mat. \cite{SM} for details).}
    \label{fig_OTOC}
\end{figure*}

The previously discussed power-law behaviors can be obtained analytically considering the stationary phase approximation (SPA) \cite{PRB_long_Kitaev_2024}. Suppl. Mat. \cite{SM} summarizes our results.

In Ref.~\cite{PRB_count_edge_2023}, the authors have obtained identical results to those shown in Fig. \ref{fig_resp_func}, but for another quantum impurity model. Their model can be obtained from Eq. (\ref{main_Hamil}) through $Jh\mu\sigma^z_1 \to Jh\sigma^z_1 + Jh\mu\sigma^x_1$, i.e., introducing an extra longitudinal boundary field of strength $Jh\mu$. Despite this distinction between the boundary terms, both models can be shown to be dual, in the sense of the Kramers-Wannier duality~\cite{KW_duality} (see Suppl. Mat. \cite{SM}). This means that the models are connected via the transformation $h\to h^{-1}$ ($J=1$). Thus, the upcoming results also apply to the model considered in~\cite{PRB_count_edge_2023} once we make $h\to h^{-1}$.

\paragraph*{Measuring the spread of correlations}
\label{sec_otoc}

We continue the analysis with a commonly used quantifier of information scrambling \cite{PRL_Quan_correl_2016,Touil2024EPL}. The so-called out-of-time order correlation (OTOC) function is given by
\begin{equation}
    \mc C(t,\mu)=\frac{1}{2}\langle |[\sigma^z_1(t),\sigma^z_1(0)]|^2 \rangle_0\,.
    \label{OTOC_main}
\end{equation}
%where $|A|^2$ signifies take $AA^{\dagger}$. 
Note that the OTOC \eqref{OTOC_main} depends on the choice of operators \cite{Touil2020QST,Tripathy2024Chaos}. Here, we have chosen a local one to emphasize once again that the local monitoring of the system can give relevant information about the nonequilibrium dynamics. Although similar to the response function (\ref{respon_fun}), it has been shown that the OTOC carries much more information about the spread of quantum correlations, even in non-chaotic systems \cite{OTOC_PRL_2017,OTOC_Kondo_PRB_2017,OTOC_Ising_PRB_2018,OTOC_inf_trap_PRB_2023,PhysRevB.107.L020202,OTOC_PRB_2024,Sur_2024,Phy_Scripta_2024}.

It is easy to see that $\mc C(t,\mu) = 1-\te{Re} \{F(t,\mu)\}$, with
\begin{equation}
F(t,\mu)=\langle \sigma^z_1(t)\sigma^z_1(0)\sigma^z_1(t)\sigma^z_1(0) \rangle_0,  
\label{OTOC_F}
\end{equation}
where the out-of-time order structure becomes evident (and $\te{Re}\{\cdot\}$ denotes the real part). In complete analogy to the response function, the short-time behavior of the OTOC is independent of the region in the $(h,\mu)$-diagram. In fact, one can show that for short times $t\ll J^{-1}$, the OTOC (\ref{OTOC_main}) grows according to, $\mc C(t\ll J^{-1},\mu)\propto t^2$, independent of $\mu$ and $h$. This behavior clearly emphasizes the slow, nonexponential spread of information. The impurity model \eqref{main_Hamil} does not support quantum chaos. 

To obtain intermediate and long-time behaviors, the calculation of the expectation value in Eq.~(\ref{OTOC_main}) is more conveniently performed introducing the real (Majorana) fermions, $a_j,b_j$, % $a_j=c^{\dagger}_j+c_j$ and $b_j=ic^{\dagger}_j-ic_j$, 
such that the spin operator at the impurity site can be rewritten as $\sigma^z_1=ia_1b_1$, and so $F(t,\mu)=\langle a_1(t)b_1(t)a_1b_1a_1(t)b_1(t)a_1b_1 \rangle_0$. As shown in Ref.~\cite{OTOC_Ising_PRB_2018}, we can cast the above expectation value in the form of the Pfaffian of a matrix whose elements are defined by the two-point correlation functions of the Majorana fermions $a_1,b_1$: $G_{rr'}(t)=\langle r(t)r'(0)\rangle_0$, $r,r'=a_1,b_1$. All the details about the calculations are relegated to the Suppl. Mat. \cite{SM}.

Figure~\ref{fig_OTOC} depicts the results for $\mc C(t,\mu)$. Regardless of the region in the $(h,\mu)$-diagram, $\mc C(t,\mu)$ reaches a maximum value around $t\approx J^{-1}$, and for $t\gtrsim J^{-1}$ it transits to another behavior. 
When $t\gg J^{-1}$, we can see how the presence of the impurity modifies the behavior of the OTOC (\ref{OTOC_main}). In the paramagnetic phase, Figs.~\ref{fig_OTOC}(a, b) show three distinct behaviors. First, without localized modes, $\mc C(t\gg J^{-1},\mu)$ decays to zero according to $t^{-3}$, like the response function, cf. Figs.~\ref{fig_resp_func}(a, b), showing a complete spread of the information as $Jt\to \infty$. 

The effects of the edge mode $\gamma_2$ on the OTOC can already be seen along the boundaries $\mu =\sqrt{1\mp 1/h}$ for $h>1$. In Figs.~\ref{fig_OTOC}(a, b), we observe that the information spreads more slowly, $\mc C(t\gg J^{-1},\mu)\sim t^{-1}$, compared to the previous situation. 

Inside the regions where only mode $\gamma_2$ contributes, $h>1$ and $\mu\lessgtr\sqrt{1\mp1/h}$, the OTOC is given by the blue curves in Figs.~\ref{fig_OTOC}(a, b). Interestingly, as time evolves, instead of a power-law decay, the OTOC remains \textit{finite} for all times. In fact, for large $t$, we have
\begin{eqnarray}
\mc C(t\gg J^{-1},\mu)\propto1-(\te{const.})\cos2\Gamma^{(2)}t,
    \label{OTOC_mode2}
\end{eqnarray}
where $\te{const.}$ is a $(h,\mu)$-dependent constant. Hence, $\mc C(t\to \infty,\mu)$ is governed by $\gamma_2$ in the regions $h>1$ and $\mu\lessgtr\sqrt{1\mp1/h}$. This long-time nonzero value of the OTOC reflects how the localized character of the wave function of the mode $\gamma_2$ around the impurity prevents the perfect spread of the information (and energy) throughout the chain. Similar results have been obtained recently for other systems showing localized edge modes \cite{OTOC_Kondo_PRB_2017,OTOC_inf_trap_PRB_2023,PhysRevB.107.L020202,Sur_2024} and for a system of interacting electrons in one dimension \cite{OTOC_PRL_2017}.

Turning to the ferromagnetic side, $h<1$, the results can be found in Fig.~\ref{fig_OTOC}(c). Due to the presence of the mode $\gamma_1$ for all $\mu$, we see that the OTOC remains finite throughout the ferromagnetic region. For $h<1$ and $\mu\leq\sqrt{1+1/h}$, $\gamma_1$ is the only localized mode that contributes to the OTOC. Since it has zero energy, $\mc C(t\gg J^{-1},\mu)$ converges to a finite value, see Fig.~\ref{fig_OTOC}(c). Finally, when $\mu>\sqrt{1+1/h}$ and $h<1$, both modes $\gamma_{1,2}$ are present, and so the OTOC shows an oscillatory long-time behavior. The corresponding analytical result reads, $\mc C(t\gg J^{-1},\mu)\propto 1-(\te{const.}^{\prime})\cos\Gamma^{(2)}t$, where $\te{const.}^{\prime}$ is another $(h,\mu)$-dependent constant. The OTOC now oscillates with frequency $\Gamma^{(2)}$ (half of the frequency found in the paramagnetic phase with a single localized mode). This is essentially the only difference from the case in which $\gamma_2$ was the only contribution. 

\paragraph*{Emergence of $X$-states}

The comparison between Figs.~\ref{fig_resp_func} and \ref{fig_OTOC} clearly shows the complementary information provided by $\Phi(t,\mu)$ and $\mc C(t,\mu)$. The nonequilibrium steady state characterized by the response and OTOC functions unambiguously exhibits the trapping of excitations in the edge modes. The natural question arises whether the emerging state has useful and interesting properties.

The time evolution of the system, starting from the ground state $|\mathbf 0\rangle$, after the perturbation $g(t)=-g_{0}\delta(t)\sigma_{1}^{z}$, is given by $\rho(t)=|\Psi(t)\rangle\langle\Psi(t)|$, where $|\Psi(t)\rangle = \left(\cos g_0\mathbf 1+i\sin g_0 e^{-iH_0t}\sigma^z_1\right)|\mathbf 0\rangle$ (see Suppl. Mat. \cite{SM}). Motivated by the correspondence between the existence of the modes $\gamma_{1,2}$, the behavior of the response and OTOC functions, and the decay of coherences induced by the perturbation between states with different occupations of localized and delocalized modes (see Suppl. Mat. \cite{SM}), we trace out the delocalized modes to obtain the reduced density operator $\rho_{\te{loc}}(t)=\Tr_{\te{del}}\rho(t),$ where $\Tr_{\te{del}}$ denotes the partial trace. 

For the remainder of the analysis, we restrict ourselves to the yellow region of Fig.~\ref{fig_system}(b), where both $\gamma_{1,2}$ modes exist. As shown in Suppl. Mat. \cite{SM}, $\rho_{\te{loc}}(t)$ then becomes
\begin{eqnarray}
\rho_{\te{loc}}(t)&=&\begin{pmatrix}\rho_{11} & 0 & 0 & \rho_{14}(t)\\
0 & \rho_{22} & \rho_{23}(t) & 0\\
0 & \rho^*_{23}(t) & \rho_{33} & 0\\
\rho^*_{14}(t) & 0 & 0 & \rho_{44}
\end{pmatrix},
\label{dens_matrix}
\end{eqnarray}
when written in the basis $\{|n_1,n_2\rangle=|0,0\rangle, |0,1\rangle, |1,0\rangle, |1,1\rangle\}$ (in this order), where $n_\ell=0,1$ is the number of excitations in the localized $\ell$ mode.

We see that $\rho_{\te{loc}}(t)$ is an $X$-state \cite{PRL_2001_Zurek,PRL_ent_death_2004,Yu_2006,PRA_Santos_2006,PRA_sud_ent_fin_tem_2008,Rau_2009,PRA_Sarandy_2009,PRA_Rau_2010,PRA_Fernan_2011,PRA_Chen_2011,PRA_Lu_2011,Quesada10092012,PRA_Huang_2013,PRA_spin_orb_X_2021}. This kind of double-qubit state describes a large and important class of entangled quantum states, ranging from pure Bell states to Werner mixed states~\cite{Rau_2009}. In the present case, the qubits consist of combinations of states having zero, one, or two fermionic excitation in the localized modes, namely, $|0,0\rangle$ and $|1,1\rangle$ for one qubit and $|0,1\rangle$ and $|1,0\rangle$ for another qubit. 

For the quantum impurity model \eqref{main_Hamil}, the $X$-structure is a consequence of the fact that $\sigma^z_1|\mathbf 0\rangle$ yields states with two or zero excitation, but also couples the localized and delocalized modes. This allows for part of the excitations to relax and hence granting access to the $4$-dimensional Hilbert space spanned by all possible occupations of the localized modes. This $X$-state form of $\rho_{\te{loc}}(t)$ cannot exist in the other regions of the $(h,\mu)$-diagram.  

\begin{figure}
    \centering
    \includegraphics[scale=0.45]{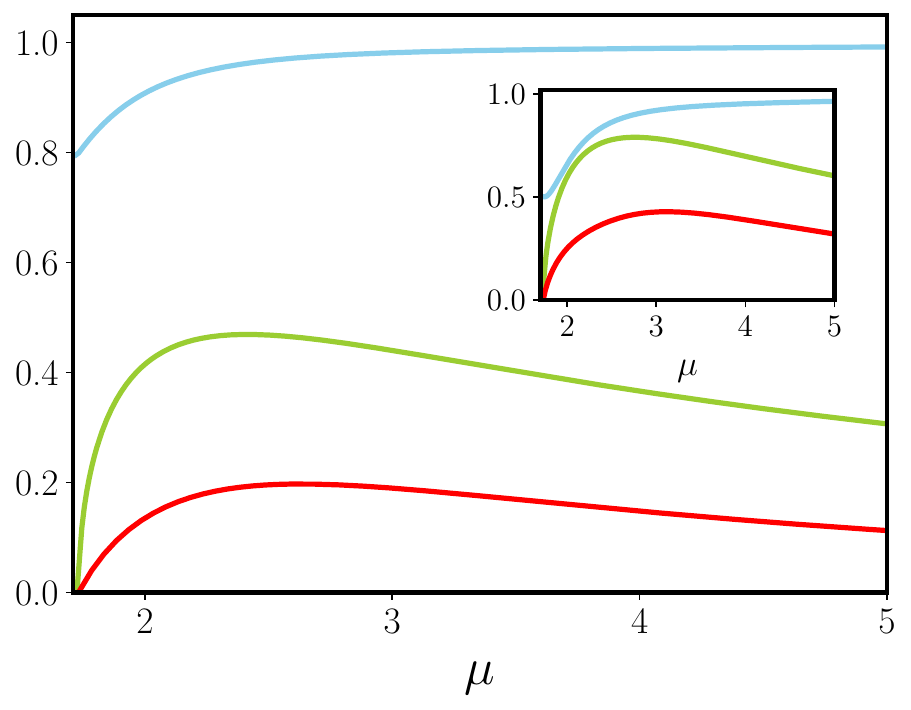}
    \caption{Purity (blue), concurrence (green), and discord (red) for the $X$-state in Eq. (\ref{dens_matrix}). We used $h=0.5$, so $\mu >1.7321$ in order to remain in the yellow region of Fig. \ref{fig_system}(b), and $g_0=0.5$. In the inset we show the same three quantities but for $g_0=\pi/2$.}
    \label{fig_Wex}
\end{figure}

In Fig.~\ref{fig_Wex} we collect information-theoretic characterizations \cite{nielsen2010quantum}, such as purity, concurrence, and discord, of $\rho_{\te{loc}}(t)$ \eqref{dens_matrix} as a function of $\mu$ for fixed values of $h$ and $g_0$ (see Suppl. Mat. \cite{SM} for other values). For $g_0 \ll 1$, the state is approximately pure, but quantum correlations, quantified by concurrence and discord, are still present. Interestingly, both concurrence and discord reach a maximal value for small but finite values of $\mu$. At the boundary $\mu=\sqrt{1+1/h}$ ($h<1$), the concurrence and discord are null due to the absence of the mode $\gamma_2$.

In the nonperturbative regime, $g_0\sim1$, we have a much more interesting situation. This is because the nontrivial part of the evolved state, $\sin g_0 e^{-iH_0t}\sigma^z_1|\mathbf 0\rangle$, which is responsible for quantum correlations, can dominate the trivial part, $\cos g_0|\mathbf 0\rangle$, when $\sin g_0>\cos g_0$. Thus, we expect to see an enhancement of quantum correlations in this regime. The inset of Fig.~\ref{fig_Wex} confirms it. Finally, for very large values of $\mu$, the quantum state becomes essentially classical.

\paragraph*{Concluding remarks}
\label{sec_conclu}

In the present analysis, we have shown the emergence of $X$-states in a quantum impurity model. This emergence is the consequence of the nonequilibrium response to a local perturbation, and it is heralded by the long-time behavior of the response function and OTOC. The state found was also shown to exhibit genuine quantum correlations. In two recent experimental works \cite{xiang_2024,liang_2024}, probing the nonequilibrium dynamics of Rydberg atomic chains, the authors were able to measure very similar OTOC as ours (\ref{OTOC_main}).

Finally, it is worth highlighting the similarity of our results and the boundary time crystals (BTC) \cite{Iemini_PRL_2018}. In our system, like in BTC, while the bulk undergoes relaxation, $\langle \sigma^z_{j\gg 1}(Jt\gg 1)\rangle= \sigma^z_{j\gg1,\te{eq.}}$, where $\sigma^z_{j\gg1,\te{eq.}}$ is the equilibrium bulk value before the perturbation, the boundary breaks time-translation symmetry, $\langle \sigma^z_{j\sim 1}(Jt\gg 1)\rangle= \varphi_j(t)$, where $\varphi_j(t)$ is a periodic function of time. A more detailed analysis of the potential emergence of BTC in impurity models is left for future work.

\acknowledgements{M.F.C. would like to thank Michael Kolodrubetz for useful discussions. This work was supported by Conselho Nacional de Desenvolvimento Cient\'{i}fico e Tecnol\'{o}gico (CNPq), Brazil, through grant No. 200267/2023-0. M.V.S.B. acknowledges the support of CNPq, under Grant No. 304120/2022-7. E.M. acknowledges the support of CNPq, under Grant No. 309584/2021-3. 
The work was also financed (M.V.S.B. and E.M.), in part, by the S\~ao Paulo Research Foundation (FAPESP), Brazil, Process Number 2022/15453-0.
S.D. and M.F.C. acknowledge support from the John Templeton Foundation under Grant No. 62422.}

\bibliography{references}

%\pagebreak
%\clearpage
\section*{Supplemental material}

In this Supplementary Material, we provide details on (i) the impurity model, (ii) the OTOC calculations, (iii) the long-time behaviors of the correlation functions, (iv) the reduced density state for the localized edge modes, and (v) the information characterization of our \texorpdfstring{$X$}{X}-state.

\section{Model}
\label{Apen_model}

The open boundary condition spin-$1/2$ chain with an edge-impurity is described by the Hamiltonian~\cite{PRB_main_ref_2016}
\begin{eqnarray}   
 H_{0}&=&-\sum_{j=1}^{N-1}\sigma^x_{j}\sigma^x_{j+1} - \:h\sum_{j=2}^{N}\sigma^z_{j}-h\mu \sigma^z_1,
\label{imp_Hamil}    
\end{eqnarray}
where $\sigma^{a=x,y,z}_j$ are the Pauli matrices at the site $j$ and $\mu$ defines the impurity at the edge. After applying the Jordan-Wigner transformation \cite{LIEB1961407,sachdev1999quantum}, $\sigma^{+}_{j}=(\sigma^x_j+i\sigma^y_j)/2= (-1)^j\exp\left(i\pi\sum_{i=1}^{j-1}c^{\dagger}_{i}c_{i}\right)c^{\dagger}_{j}$ and $\sigma^z_{j}= 2c^{\dagger}_{j}c_{j}-1$, the  Hamiltonian of Eq.~\eqref{imp_Hamil} reads
\begin{equation}
    H_0=\sum_{ij}\left[c^{\dagger}_iA_{ij}c_j+\frac{1}{2}\left(c^{\dagger}_iB_{ij}c^{\dagger}_j +\te{H.c.}\right)\right],
    \label{imp_Hamil_2} 
\end{equation}
where $A_{ij}=2h\left[(1-\mu)\delta_{i1}-1\right]\delta_{ij}+ \left(\delta_{i,j+1}+\delta_{i+1,j}\right)=A_{ji}$ and $B_{ij}=(\delta_{i+1,j}-\delta_{i,j+1})=-B_{ji}$. Following \cite{LIEB1961407,Losonczi1992-cl,Tridi_matr_Yueh2005,PRB_main_ref_2016}, the Hamiltonian (\ref{imp_Hamil_2}) can be brought to the form
\begin{eqnarray}
H_0&=&\sum_{\kappa}\Gamma_{\kappa}\gamma^{\dagger}_{\kappa}\gamma_{\kappa} + \te{const.},
    \label{imp_Hamil_3}
\end{eqnarray}
where $c_j = \sum_{\kappa}u_{j\kappa}\gamma_{\kappa}+v_{j\kappa}\gamma^{\dagger}_{\kappa}$ represents the Bogoliubov transformation \cite{LIEB1961407}. $\gamma_{\kappa}$ are fermionic degrees of freedom: $\gamma_{1,2}$ for the localized modes, and $\gamma_{k}$ for the delocalized ones. To guarantee the correct fermionic anticommutation relations, $\{\gamma^{\dagger}_{\kappa},\gamma_{\kappa'}\}=\delta_{\kappa\kappa'}$ and $\{\gamma^{\dagger},\gamma_{\kappa'}\}=0$, the wave functions $u_{j\kappa}$ and $v_{j\kappa}$ need to satisfy (in matrix notation): $\mathbf 1=uu^{\te{T}}+vv^{\te{T}}$ and $\mathbf 0=uv^{\te{T}}+vu^{\te{T}}$, where $\mathbf 1$ is the identity matrix, $\mathbf 0$ is the null matrix, and $\te{T}$ denotes the transpose operation. The inverse transformation reads $\gamma_{\kappa}=\sum_{j}u_{j\kappa}c_{j}+v_{j\kappa}c^{\dagger}_{j}$. In what follows, we will work with the quantities $\psi_{j\kappa }\equiv u_{j\kappa }-v_{j\kappa }$ and $\phi_{j\kappa }\equiv u_{j\kappa }+v_{j\kappa }$.

\begin{table*}
\scalebox{1.0}{
\begin{tabular}{|c|c|c|}
	\hline
	Region & $\Phi(Jt\gg 1,\mu)$  & $\mc C(Jt\gg 1,\mu)$ \\
	\hline\hline
	$h>1$ and $\mu\lessgtr\sqrt{1\mp1/h}$ & $t^{-3/2}$ & Eq. (9) \\
	\hline
	$h>1$ and $\mu=\sqrt{1\mp1/h}$ & $t^{-2}$ & $t^{-1}$ \\
	\hline
	$h>1$ and $\sqrt{1-1/h}<\mu<\sqrt{1+1/h}$ & $t^{-3}$ & $t^{-3}$ \\
	\hline
	$h<1$ and $\mu<\sqrt{1+1/h}$ & $t^{-3/2}$ & $\frac{2(1-h^2)^2}{[1+h^2(\mu^2-1)]^2}+\mc O(t^{-3/2})$ \\
	\hline
	$h<1$ and $\mu=\sqrt{1+1/h}$ & $t^{-1/2}$ & $2(1-h)^2+\mc O(t^{-1/2})$ \\
	\hline
	$h<1$ and $\mu>\sqrt{1+1/h}$ & Eq. (6) & $\propto 1-(\te{const.}^{\prime})\cos\Gamma^{(2)}t$ \\
	\hline
\end{tabular}}
\caption{Asymptotic behaviors of the response function, Eq. (4), and the OTOC, Eq. (7), for the different boundary phases of the model in Eq.~(\ref{imp_Hamil}).}
\label{t1}
\end{table*}

As discussed in the main text, the edge modes $\gamma_{\ell=1,2}$ only appear in specific regions of the $(h,\mu)$-phase diagram (see Fig. 1(b) in the main text). For the  $\gamma_{2}$ mode, these regions are: $|\mu|>\sqrt{1+1/h}$ for all $h>0$ and $|\mu|<\sqrt{1-1/h}$ for $h>1$. The energy and wave functions of this mode are,
\begin{eqnarray}
    \Gamma^{(2)}&=&2|\mu|\sqrt{\frac{1+(\mu^2-1)h^2}{\mu^2-1}},\nonumber\\
    \psi^{(2)}_j&=&\frac{(-1)^jh^{-j}\sqrt{(\mu^2-1)^2h^2-1}}{(\mu^2-1)^j},\nonumber\\
    \phi^{(2)}_j&=&\frac{1}{\Gamma^{(2)}}\big[-2h\psi^{(2)}_j + 2h(1-\mu)\delta_{j1}\psi^{(2)}_1\nonumber\\
    &&+2(1-\delta_{j1})\psi^{(2)}_{j-1}\big],
\end{eqnarray}
where, hereafter, we use the superscript $(\ell)$ to label the mode $\ell=1,2$. The  $\gamma_1$ mode exists in the entire region $h<1$ (topological phase of the Kitaev chain \cite{Kitaev_2001}). For this mode, we have
\begin{eqnarray}
    \Gamma^{(1)}&\simeq&\frac{2|\mu|(1-h^2)h^N}{\sqrt{|1+h^2(\mu^2-1)|}},\nonumber\\
    \psi^{(1)}_j&=&\sqrt{1-h^2}h^N\left(h^{-j}-\frac{\mu^2h^j}{1+h^2(\mu^2-1)}\right),\nonumber\\
    \phi^{(1)}_j&=&-\frac{\sgn(\mu)\sqrt{1-h^2}\sqrt{|1+h^2(\mu^2-1)|}}{h[1+h^2(\mu^2-1)]}\nonumber\\
    &&\times\left[\delta_{j1}(1-\mu)+\mu\right]h^j. 
\end{eqnarray}
where $\sgn\{\cdot\}$ is the sign function. Notice that in the limit $N\to \infty$, $\Gamma^{(1)},\psi^{(1)}_1\to 0$.

Finally, for the delocalized modes $\gamma_k$, we have
\begin{eqnarray}
    \Gamma_k&=&2\sqrt{1+h^2-2h\cos k},\nonumber\\
    \psi_{jk}&=&\sqrt{\frac{2}{N}}\frac{\sin jk+h(\mu^2-1)\sin (j-1)k}{\sqrt{1+h^2(\mu^2-1)^2+2h(\mu^2-1)\cos k}},\nonumber\\
    \phi_{jk}&=&\frac{1}{\Gamma_k}\big[-2h\psi_{jk}+2h(1-\mu)\psi_{1k}\delta_{j1}\nonumber\\
    &&+2(1-\delta_{j1})\psi_{j-1,k}\big].
    \label{bulk}
\end{eqnarray}
In the continuum limit, we can take $k \in [0,\pi]$ as usual \cite{PRB_main_ref_2016}. We see that the system gap closes at $h=1$, independently of $\mu$, where the system  undergoes a quantum phase transition~\cite{sachdev1999quantum}.

\subsection{Majorana representation}

The model of Eq.~(\ref{imp_Hamil}) can be rewritten in terms of Majorana fermions. As discussed in the main text, this representation is useful for the calculation of the OTOC.

Decomposing the spinless fermionic operators $c_i$ in the basis, $c^{\dagger}_j=\frac{a_j - ib_j}{2}$ and $c_j=\frac{a_j+ib_j}{2}$,
%\begin{eqnarray}
%    c^{\dagger}_j&=&\frac{a_j - ib_j}{2},\\
%    c_j&=&\frac{a_j+ib_j}{2},
%\end{eqnarray}
where $a_j$ and $b_j$ are Majorana fermions, $a^2_j=b^2_j=1$, $\{a_j,a_{j'}\}=\{b_j,b_{j'}\}=2\delta_{jj'}$ and $\{a_j,b_{j'}\}=0$, the Hamiltonian of Eq.~(\ref{imp_Hamil_2}) is recast as
\begin{equation}
    H_0=ih\mu b_1a_1+ih\sum_{j=2}^{N}b_ja_{j}-i\sum_{j=1}^{N-1}b_ja_{j+1}+\te{const.}.
\end{equation}
In particular, when $\mu=0$, we can see that the Majorana fermion $a_1$ totally decouples from the rest of the chain, $[a_1,H_0]=0$, and thus it becomes a Majorana zero mode. 

Another Majorana representation for the model (\ref{imp_Hamil}) can be obtained by defining the new real fermions as follows: $\eta_{A,j}=i(c^{\dagger}_j-c_j)$ and $\eta_{B,j-1}=(c^{\dagger}_j+c_j)$, where $\eta^2_{A,j}=\eta^2_{B,j}=1$, $\{\eta_{A,j},\eta_{A,j'}\}=\{\eta_{B,j},\eta_{B,j'}\}=2\delta_{jj'}$ and $\{\eta_{A,j},\eta_{B,j'}\}=0$. Using this new basis, the Hamiltonian (\ref{imp_Hamil}) reads now
\begin{equation}
    h^{-1}H_0=-i\mu \eta_{B,0}\eta_{A,1}-i\sum_{j=1}^{N-1}\eta_{B,j}\eta_{A,j+1}-ih^{-1}\sum_{j=1}^{N}\eta_{A,j}\eta_{B,j},
\end{equation}
where the const. term was ignored. The right-hand side of the above equation is exactly the Hamiltonian considered in \cite{PRB_count_edge_2023} (see its Eq. (4)) once we make $h^{-1}\to h$ and identify $\eta_{B,0}$ as the ancilla Majorana fermion defined there.

\section{OTOC calculation}
\label{Apen_OTOC}
Here, we are interested in the OTOC,
\begin{eqnarray}
    \mc C(t,\mu)&=&\frac{1}{2}\langle |[\sigma^z_1(t),\sigma^z_1(0)]|^2 \rangle_0,\nonumber\\
    &=&1-\te{Re} \{F(t,\mu)\},
\end{eqnarray}
where $F(t,\mu)=\langle \sigma^z_1(t)\sigma^z_1(0)\sigma^z_1(t)\sigma^z_1(0) \rangle_0$ and $\te{Re}\{\cdots\}$ denotes the real part.
Using the first Majorana representation discussed earlier,
%\begin{eqnarray}
%    a_n&=&c^{\dagger}_n + c_n,\\
%    b_n&=&i(c^{\dagger}_n-c_n),
%\end{eqnarray}
we can rewrite $\sigma^z_1=ia_1b_1$ so that
\begin{equation}
    F(t,\mu)=\langle a_1(t)b_1(t)a_1b_1a_1(t)b_1(t)a_1b_1 \rangle_0.
\end{equation}
The above OTOC can be calculated applying Wick's theorem since $a_1$ and $b_1$ are linear combinations of $\gamma_{\kappa}$ and $\gamma^{\dagger}_{\kappa}$. However, as shown in reference \cite{OTOC_Ising_PRB_2018}, this expectation value can be cast in the form of a Pfaffian of a matrix $\mc M(t)$,
\begin{equation}
    F(t,\mu)=\Pf\{\mc M(t)\}.
\end{equation}
For our case,
\begin{equation}
\mc M(t) = \begin{pmatrix}
\mc A(0) & \mc B(t) & \mathbf 1 +\mc A(0) & \mc B(t)\\
-\mc B^{\te{T}}(t)& \mc A(0) & \mc B(-t) & \mathbf 1+\mc A(0)\\
-\mathbf 1+\mc A(0) & -\mc B^{\te{T}}(-t) & \mc A(0) & \mc B(t)\\
-\mc B^{\te{T}}(t) & -\mathbf 1+\mc A(0) & -\mc B^{\te{T}}(t) & \mc A(0)
\end{pmatrix},
\end{equation}
where we defined the $2\times 2$ matrices
\begin{eqnarray}
\mc A(t)&=&\begin{pmatrix}
0 & G_{ab}(t)\\
-G_{ab}(t) & 0
\end{pmatrix},\\
\mc B(t)&=&\begin{pmatrix}
G_{aa}(t) & G_{ab}(t)\\
G_{ba}(t) & G_{bb}(t)
\end{pmatrix}.
\end{eqnarray}
Here, $G_{rr'}(t)\equiv\langle r(t)r'(0)  \rangle_0$, where $r,r'=a_1,b_1$, and $\mathbf 1$ is the $2\times 2$ identity matrix. Notice that $\mc {A}(t)=-\mc A^\te{T}(t)$.

Since, like the determinant, the Pfaffian is invariant under the addition of rows and columns, we obtain
\begin{equation}
    F(t,\mu)=\Pf\{\Tilde{\mc M}(t)\},
\end{equation}
where
\begin{equation}
\Tilde{\mc M}(t) = \begin{pmatrix}
\mc A(0) & \mc B(t) & \mathbf 1 & \mathbf0\\
-\mc B^{\te{T}}(t)& \mc A(0) & \mathbf0 & \mathbf 1\\
-\mathbf 1 & \mathbf0 & \mathbf0 & \mc B(t)+\mc B^{\te{T}}(-t)\\
\mathbf0 & -\mathbf 1 & -\mc B^{\te{T}}(t)-\mc B(-t) & \mathbf0
\end{pmatrix},
\end{equation}
and $\mathbf 0$ is the $2\times 2$ null matrix. Performing the calculation, we find~\cite{OTOC_Ising_PRB_2018}
\begin{widetext}
\begin{eqnarray}
    2^{-1}\mc C(t,\mu)&=&(\te {Re}\{G_{ba}(t)\})^2+(\te {Re}\{G_{ab}(t)\})^2+(\te {Re}\{G_{aa}(t)\})^2+(\te {Re}\{G_{bb}(t)\})^2\nonumber\\
    &&-2\Big[\te {Re}\{G_{ba}(t)\}\te {Re}\{G_{ab}(t)\}-\te {Re}\{G_{aa}(t)\}\te {Re}\{G_{bb}(t)\} \Big]\Big[\te {Re}\{G^2_{ab}(0)\}\nonumber\\ 
    &&+\te {Re}\{G_{ab}(t)\}\te {Re}\{G_{ba}(t)\}-\te {Im}\{G_{ab}(t)\}\te {Im}\{G_{ba}(t)\}\nonumber\\
    &&-\te {Re}\{G_{aa}(t)\}\te {Re}\{G_{bb}(t)\}+\te {Im}\{G_{aa}(t)\}\te {Im}\{G_{bb}(t)\}\Big],
    \label{OTOC_full_expr}
\end{eqnarray}
\end{widetext}
where $\te{Im}\{\cdot\}$ denotes the imaginary part.

The two-point Majorana correlation functions in the above expression, $G_{rr'}(t)$, are given by
\begin{eqnarray}
    G_{ab}(t)&=&i\sum_{\kappa}\phi_{\kappa}\psi_{\kappa}e^{-i\Gamma_{\kappa}t},\nonumber\\
    G_{ba}(t)&=&[G_{ab}(-t)]^*,\nonumber\\
    G_{aa}(t)&=&\sum_{\kappa}\phi^2_{\kappa}e^{-i\Gamma_{\kappa}t},\nonumber\\
    G_{bb}(t)&=&\sum_{\kappa}\psi^2_{\kappa}e^{-i\Gamma_{\kappa}t},
    \label{Maj_corr_func}
\end{eqnarray}
where the above summations include both the delocalized and the localized modes, and we used the notation $\psi_{1\kappa}\equiv \psi_{\kappa}$. %Using the stationary phase approximation (SPA) discussed in the main text, we can obtain the asymptotic long-time behavior of $G_{ij}(t)$.

\section{Stationary phase approximation (SPA)}
\label{Apen_SPA}
The long-time power-law behaviors observed for the response function and the OTOC (see Figs. 2 and 3  in the main text) can be understood by applying the stationary phase approximation (SPA) \cite{PRB_long_Kitaev_2024}. 

\subsection{Response function $\Phi(t,\mu)$}

In the limit $N\to\infty$, the summation over the bulk modes in Eq. (5) is converted to an integral. In the red region of Fig. 1(b), where we only have delocalized modes, the response function is given by
\begin{eqnarray}
    \Phi(t,\mu)&=&\int_0^{\pi}dkdk'f_{kk'}\sin[(\Gamma_{k}+\Gamma_{k'})t].
    \label{respon_fun_3}
\end{eqnarray}
The long-time limit ($t\gg J^{-1}$)  of 
$\Phi(t,\mu)$ is obtained by performing the integration of $f_{kk'}\sin[(\Gamma_{k}+\Gamma_{k'})t]$ expanded near the extreme points of $\Gamma_{k}+\Gamma_{k'}$: $\{(0,0),(0,\pi),(\pi,0),(\pi,\pi)\}$ [see Eq.~(\ref{bulk})]. Expanding around the points $\{(0,0),(\pi,\pi)\}$ gives us $f_{kk'}\sim k^2k'^2(k^2-k'^2)$, which produces a  decay faster than $t^{-3}$. If  we expand $f_{kk'}$ around $(0,\pi)$ (or, equivalently, around $(\pi,0)$), we  obtain $f_{kk'}\sim k^2(k'-\pi)^2$. This leads to, $\Phi(t\gg J^{-1},\mu)\sim \left(t^{-3/2}\right)^2=t^{-3}$, which agrees with the numerical calculation of Eq. (5), see Fig. 2(a, b) in the main text. 

When only one of the localized modes $\gamma_{\ell=1,2}$ is present (and far from the boundaries), besides the contribution in Eq. (\ref{respon_fun_3}), we also have the term $\int_0^{\pi}dk\:(f_{k\ell}+f_{\ell k})\sin[(\Gamma_{k}+\Gamma^{(\ell)})t]$. For the latter, obviously, the extreme points are $k=0,\pi$, and both of them give the same result, $f_{k\ell}\sim k^2$ (when expanded around $k=\pi$, we have done a $\pi$-translation). Thus, $\Phi(t\gg J^{-1}, \mu)\sim t^{-3/2}$. At the boundaries, $h>1$ and $\mu=\sqrt{1\mp1/h}$, $f_{k k'}$ behaves as $f_{kk'}\sim k^2$ around $(0,\pi)$. Then, $\Phi(t\gg J^{-1}, \mu)\sim t^{-3/2}t^{-1/2}=t^{-2}$. However, at the boundary $\mu=\sqrt{1+ 1/h}$ and $h<1$, the mode $\gamma_1$ contributes, making $f_{k,\ell=1}\sim 1$ around $k=\pi$. This produces the slower decay seen in Fig. 2(c), $\Phi(t\gg J^{-1}, \mu)\sim t^{-1/2}$.

\subsection{OTOC $\mc C(t,\mu)$}

As we saw in the Sec. \ref{Apen_OTOC}, to calculate $\mc C(t,\mu)$ we need to know the two-point Majorana correlation functions, $G_{rr'}(t)$, see Eq. (\ref{OTOC_full_expr}). The long-time behavior of $G_{rr'}(t)$ can  be extracted from the SPA discussed above. In the red region of Fig. 1(b) (only delocalized modes), from Eq. (\ref{bulk}) we obtain $\psi_k,\phi_k\sim k$ around $k=0$. Thus, $G_{rr'}(t\gg J^{-1})\sim t^{-3/2}$ (see Eq. (\ref{Maj_corr_func})). This leads to $\mc C(t\gg J^{-1},\mu)\sim t^{-3}$. At the boundaries $h>1$ and $\mu=\sqrt{1\mp1/h}$, $\psi_k,\phi_k\sim 1$ (for $\mu=\sqrt{1-1/h}$ we have expanded around $k=0$ while for $\mu=\sqrt{1+1/h}$ around $k=\pi$). Then, $G_{rr'}(t\gg J^{-1})\sim t^{-1/2}$, which produces $\mc C(t\gg J^{-1},\mu)\sim t^{-1}$.

In table \ref{t1} we summarize all the results for the asymptotic behaviors of $\Phi(t,\mu)$ and $\mc C(t,\mu)$.

\section{Reduced state for the localized modes}
\label{Apen_X_states}

For the system initially prepared in its ground state $|\mathbf 0\rangle$ within the yellow region of Fig. 1(b) (where both $\gamma_{1,2}$ edge modes are present), we turn on the impulsive perturbation $-g_0\delta(t)\sigma^z_1$. The system state at time $t>0$ will be given by
\begin{eqnarray}
    |\Psi(t)\rangle = e^{-iH_0t}T\exp\left[-i\int_0^t dt' \delta H(t') \right]|\mathbf 0\rangle,
\end{eqnarray}
where $\delta H(t)=-g_0\delta(t)\sigma^z_1(t)$, with $\sigma^z_1(t)=e^{iH_0t}\sigma^z_1e^{-iH_0t}$, and $T$ is the time ordering operator. Using the Dirac delta function property,
\begin{eqnarray}
    |\Psi(t)\rangle&=&\left(\cos g_0\mathbf 1+i\sin g_0 e^{-iH_0t}\sigma^z_1\right)|\mathbf 0\rangle,
\end{eqnarray}
where it was considered that $H_0|\mathbf 0\rangle=0$. Thus, the system density operator reads,
\begin{eqnarray}
    \rho(t)&=&|\Psi(t)\rangle\langle\Psi(t)|,\nonumber\\
    &=&\cos^2g_0\left(|\mathbf 0\rangle\langle\mathbf 0|\right) -i\frac{\sin2g_0}{2}\left(|\mathbf 0\rangle\langle\mathbf 0|\sigma^z_1e^{iH_0t}\right),\nonumber\\
    &&+i\frac{\sin2g_0}{2}\left(e^{-iH_0t}\sigma^z_1|\mathbf 0\rangle\langle\mathbf 0|\right),\nonumber\\
    &&+\sin^2g_0\left(e^{-iH_0t}\sigma^z_1|\mathbf 0\rangle\langle\mathbf 0|\sigma^z_1e^{iH_0t}\right).
    \label{rho_tot}
\end{eqnarray}

The density operator $\rho(t)$, being the state of the full system, provides the time evolution of any average value, in particular, the local magnetization of the impurity, given by $\langle\sigma^z_1(t)\rangle$. However, $\langle \sigma^z_1(t)\rangle$ can also be obtained from the response function (4) in the perturbative regime and, as shown in Fig.2(c) of the main text, its long-time behavior has persistent oscillations when the two localized modes are present. Next, we show that this is a signature of the $X$-state in the reduced density operator of the localized modes and that couplings (in the sense of non-diagonal matrix elements of $\rho(t)$) introduced by the perturbation between the edge and bulk modes with different number of excitations in the later ones decay in the long-time limit. To see this, we express $\langle\sigma^z_1(t)\rangle$ as follows,
\begin{eqnarray}
    \langle \sigma^z_1(t)\rangle&=&\Tr\{\rho(t)\sigma^z_1\},\nonumber\\
    &=&\sum_{n,n'}\langle n | \rho(t)|n'\rangle\langle n'|\sigma^z_1 |n\rangle,
    \label{sz_imp}
\end{eqnarray}
where $|n\rangle$ and $|n'\rangle$ are two Fock states: $|n\rangle\equiv |n_1,n_2\rangle\otimes|\nu_1,\nu_2,\cdots\rangle$, being $|n_1,n_2\rangle$ (with $n_{\ell=1,2}=0,1$) the localized edge modes part and $|\nu_1,\nu_2,\cdots\rangle$ (with $\nu_{i}=0,1$) the delocalized modes part. Because $\sigma^z_1$ creates zero or two $\gamma_{\kappa}$ excitations ($\sigma^z_{j}= 2c^{\dagger}_{j}c_{j}-1$), the non-zero terms in the above expression are terms where $|n\rangle$ and $|n'\rangle$ differ by zero or two excitations. These excitations can occupy edge or bulk modes. However, in the long-time limit, $t\gg J^{-1}$, not all of those terms will survive due to the presence of high oscillatory factors. For instance, let us analyze the matrix elements $\langle kk'|\otimes\langle0,0|\rho(t)|1,0\rangle\otimes|k''\rangle$, where $|k\rangle=\gamma^{\dagger}_k|0_d\rangle$ and $|kk'\rangle=\gamma^{\dagger}_k\gamma^{\dagger}_{k'}|0_d\rangle$, being $|0_d\rangle$ the delocalized fermionic vacuum, $\nu_i=0$. The last term of Eq. (\ref{rho_tot}) gives us (the others are zero), 
\begin{eqnarray}
    \langle kk'|\otimes\langle0,0|\rho(t)|1,0\rangle\otimes|k''\rangle&\sim&e^{-i(\Gamma_k+\Gamma_{k'}-\Gamma_{k''})t}.
\end{eqnarray}
Thus, the sum over all states like this in Eq. (\ref{sz_imp}), is a sum of high oscillatory terms, which goes to zero when $t\gg J^{-1}$. Therefore, the only surviving coherences in the long-time limit are those corresponding to Fock states with equal number of excitations in the bulk modes. In the long-time limit, the relevant content of $\rho(t)$ is hence concentrated in the subspace spanned by the states with 0, 1 and 2 excitations in the edge modes, i.e., in the reduced density operator obtained by
%Therefore, the stationary state of the system is obtained by tracing out these terms, or, more precisely, by tracing out the delocalized modes. This proceed gives us the reduced density state for the localized edge modes,
\begin{eqnarray}
    \rho_{\te{loc}}(t)&=&\Tr_{\te{del}}\rho(t),\nonumber\\
    &=&\sum_{\nu_1,\nu_2,\cdots}\langle \nu_1,\nu_2,\cdots|\rho(t)|\nu_1,\nu_2,\cdots\rangle.
\end{eqnarray}
%where $\nu_i=0,1$ and $|\nu_1,\nu_2,\cdots\rangle$ is a Fock state for a given number of bulk excitations $\nu_1+\nu_2+\cdots$.  

After performing the above calculation, we find
\begin{widetext}
\begin{eqnarray}
    \rho_{\te{loc}}(t)&=&\rho_{11}|0,0\rangle\langle0,0|+\rho_{22}|0,1\rangle\langle0,1|+\rho_{33}|1,0\rangle\langle1,0|+\rho_{44}|1,1\rangle\langle
    1,1|,\nonumber\\
    &&+\rho_{14}(t)|0,0\rangle\langle1,1|+\rho^*_{14}(t)|1,1\rangle\langle0,0|+\rho^*_{23}(t)
|1,0\rangle\langle0,1|+\rho_{23}(t)
|0,1\rangle\langle1,0|.
\end{eqnarray}
\end{widetext}
In the matrix representation this reads,
\begin{eqnarray}
\rho_{\te{loc}}(t)&=&\begin{pmatrix}
\rho_{11} & 0 & 0 & \rho_{14}(t)\\
0 & \rho_{22} & \rho_{23}(t) & 0\\
0 & \rho^*_{23}(t) & \rho_{33} & 0\\
\rho^*_{14}(t) & 0 & 0 & \rho_{44}
\end{pmatrix},
\end{eqnarray}
where
%\begin{widetext}
\begin{eqnarray}
\rho_{11}&=&\cos^2g_0+\sin^2g_0\left(|\Upsilon_1|^2+\sum_{kk' (k\neq k)}|\Upsilon_2(k,k')|^2\right),\nonumber
\end{eqnarray}
\begin{eqnarray}
    \rho_{22}&=&\sin^2g_0\sum_{k}|\Upsilon_3(k)|^2,\nonumber\\
    \rho_{33}&=&\sin^2g_0\sum_{k}|\Upsilon_4(k)|^2,\nonumber\\
    \rho_{44}&=&\sin^2g_0|\Upsilon_5|^2,\nonumber\\
    \rho_{14}(t)&=&\Upsilon^*_5\left(\Upsilon_1\sin^2g_0-i\frac{\sin2g_0}{2}\right),\nonumber\\
    \rho_{23}(t)&=&\sin^2g_0\sum_{k}\Upsilon_3(k)\Upsilon^*_4(k).
\end{eqnarray}
The functions $\Upsilon_i$ in the above expressions are
\begin{eqnarray}
    \Upsilon_1&=&-1+2\sum_{k}v^2_{k},\\
    \Upsilon_2(k,k')&=&2e^{-i(\Gamma_k+\Gamma_{k'})t}v_{k'} u_{k},\\
    \Upsilon_3(k)&=&2e^{-i(\Gamma_k+\Gamma^{(2)})t}\left[v^{(2)}u_{k}-v_ku^{(2)}\right],\\
    \Upsilon_4(k)&=&2e^{-i(\Gamma_k+\Gamma^{(1)})t}\left[v^{(1)}u_{k}-v_ku^{(1)}\right],\\
    \Upsilon_5&=&2e^{-i(\Gamma^{(2)}+\Gamma^{(1)})t}\left[v^{(2)}u^{(1)}-v^{(1)}u^{(2)}\right].
\end{eqnarray}
Notice that $\Upsilon_1$ is time-independent.

\section{Purity, entanglement and discord}\label{App.entangled}

\begin{figure*}
    \centering
    \includegraphics[width=0.8\linewidth]{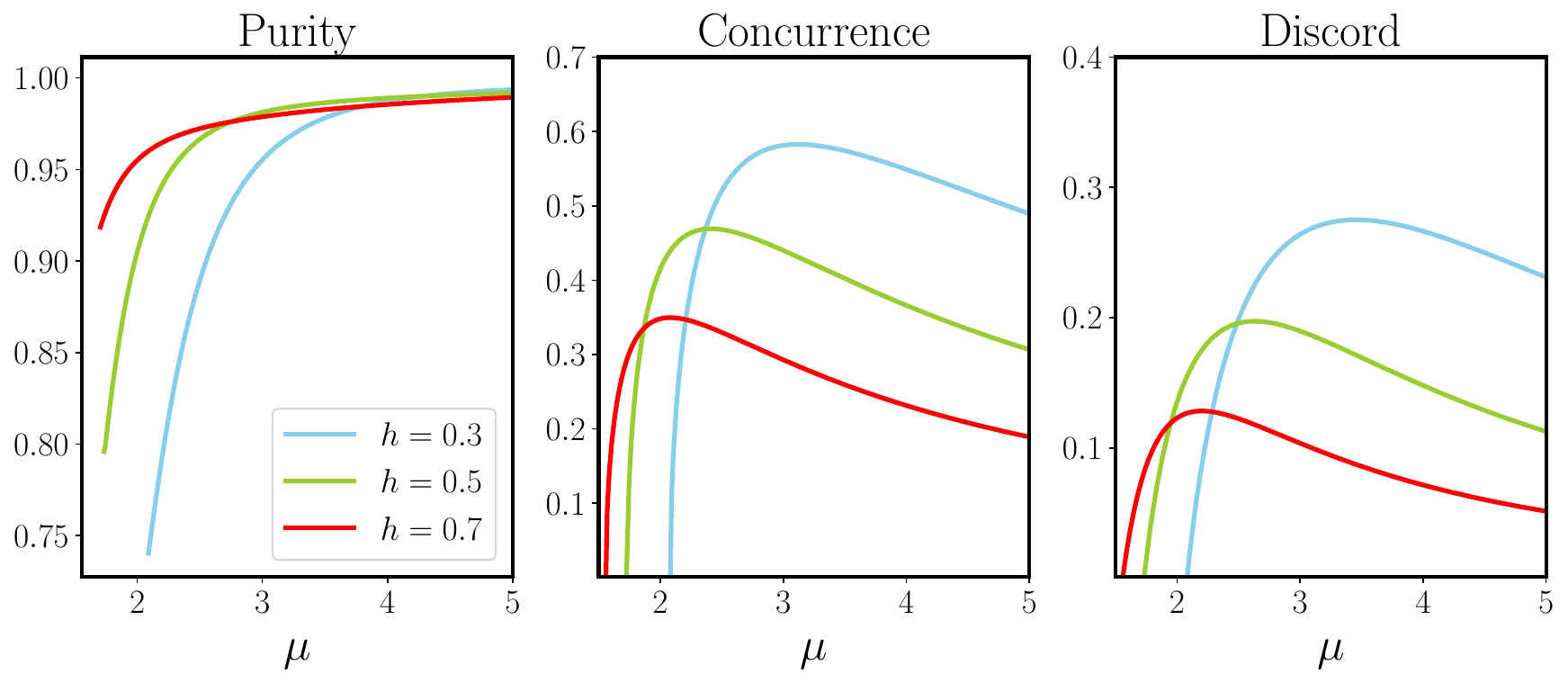}
    \caption{Purity, concurrence and discord for the $X$-state in Eq. (9). The lower $\mu$ cutoffs of the purity curves come from the need to satisfy $\mu >\sqrt{1+1/h}$. Here we take $g_0=0.5$. }
    \label{fig_PCD}
\end{figure*}

\begin{figure*}
    \centering
    \includegraphics[width=0.8\linewidth]{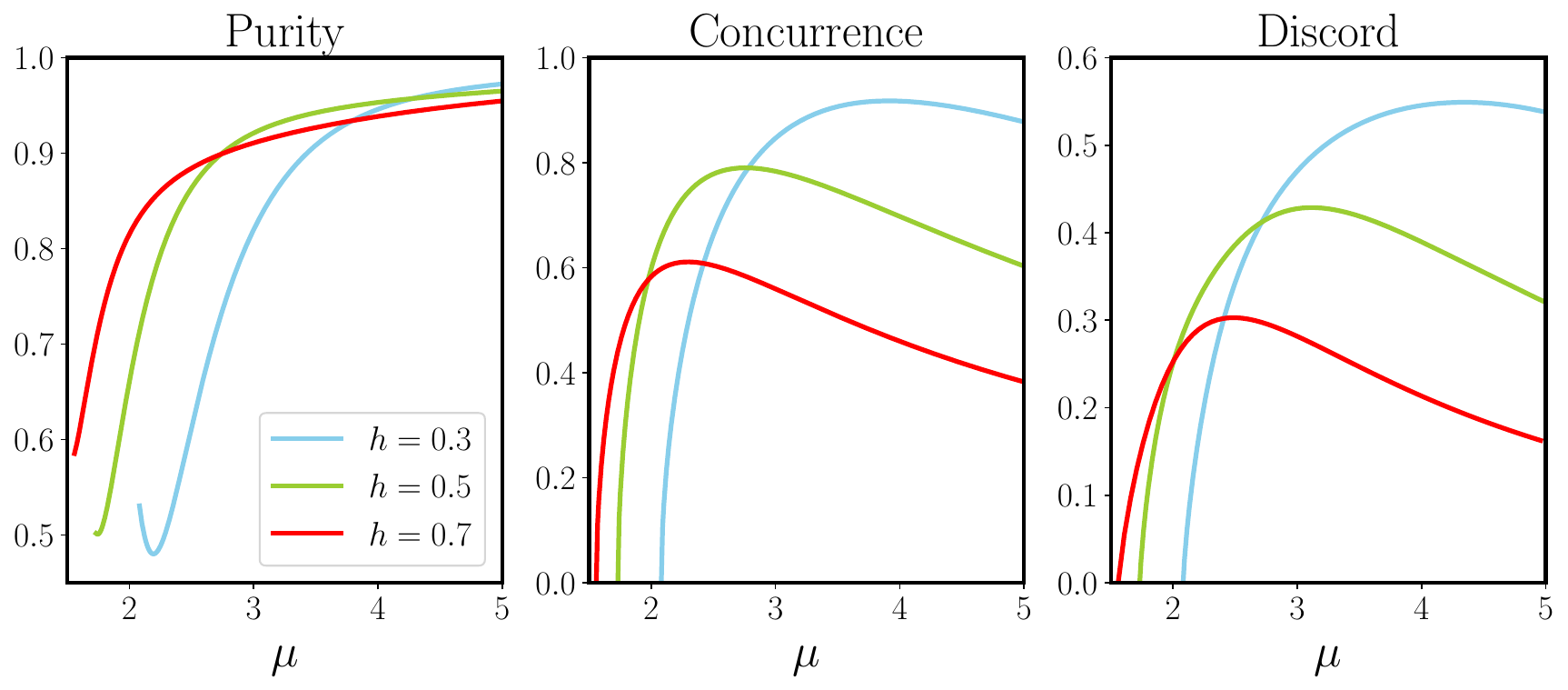}
    \caption{Purity, concurrence and discord for the $X$-state in Eq. (9). The lower $\mu$ cutoffs of the purity curves come from the need to satisfy $\mu >\sqrt{1+1/h}$. Here we take $g_0=\pi/2$.}
    \label{fig_PCD2}
\end{figure*}

The purity is obtained as usual from $\te{Tr}_{\te{loc}} \rho_{\te{loc}}^{2}(t)$, with $\rho_{\te{loc}}(t)$ given by Eq.~(10). The entanglement measure is given by the concurrence, $0<C(\rho_{\te{loc}}(t))<1$, which can be obtained in terms of the matrix elements of the $X$-state as follows \cite{Yu_2006}
\begin{eqnarray}
    \lefteqn{C(\rho_{\te{loc}}(t)) =}\nonumber\\
    &&\te{max}\{0,2(|\rho_{14}(t)|-\sqrt{\rho_{22}\rho_{33}}),2(|\rho_{23}(t)|-\sqrt{\rho_{11}\rho_{44}})\}.\nonumber\\
    \label{concurrence}
\end{eqnarray}
Finally, the discord is obtained following Refs.~\cite{PRA_Rau_2010,PRA_Huang_2013}. In Figs.~\ref{fig_PCD} and \ref{fig_PCD2} we show the results for these three quantities assuming different $h$ values.

\end{document}